# Scalable Density-Based Distributed Clustering

Eshref Januzaj[1], Hans-Peter Kriegel[2], Martin Pfeifle[2]

[1]Braunschweig University of Technology, Software Systems Engineering
http://www.sse.cs.tu-bs.de, januzaj@sse.cs.tu-bs.de
[2]University of Munich, Institute for Computer Science
http://www.dbs.ifi.lmu.de, {kriegel,pfeifle}@dbs.ifi.lmu.de

**Abstract.** Clustering has become an increasingly important task in analysing huge amounts of data. Traditional applications require that all data has to be located at the site where it is scrutinized. Nowadays, large amounts of heterogeneous, complex data reside on different, independently working computers which are connected to each other via local or wide area networks. In this paper, we propose a scalable density-based distributed clustering algorithm which allows a user-defined trade-off between clustering quality and the number of transmitted objects from the different local sites to a global server site. Our approach consists of the following steps: First, we order all objects located at a local site according to a quality criterion reflecting their suitability to serve as local representatives. Then we send the best of these representatives to a server site where they are clustered with a slightly enhanced density-based clustering algorithm. This approach is very efficient, because the local determination of suitable representatives can be carried out quickly and independently from each other. Furthermore, based on the scalable number of the most suitable local representatives, the global clustering can be done very effectively and efficiently. In our experimental evaluation, we will show that our new scalable density-based distributed clustering approach results in high quality clusterings with scalable transmission cost.

## 1 Introduction

Density-based clustering has proven to be very effective for analyzing large amounts of heterogeneous, complex data, e.g. for clustering of complex objects [1][4], for clustering of multi-represented objects [9], and for visually mining through cluster hierarchies [2]. All these approaches require full access to the data which is going to be analyzed, i.e. the data has to be located at one single site. Nowadays, large amounts of heterogeneous, complex data reside on different, independently working computers which are connected to each other via local or wide area networks (LANs or WANs). Examples comprise distributed mobile networks, sensor networks or supermarket chains where check-out scanners, located at different stores, gather data unremittingly. Furthermore, international companies such as DaimlerChrysler have some data which is located in Europe and some data in the US. Those companies have various reasons why the data cannot be transmitted to a central site, e.g. limited bandwidth or security aspects. Another example is WAL-MART featuring the largest civil database in the world, consisting of more than 200 terabytes of data [11]. Every night all data is transmitted to Betonville from the different stores via the largest privately hold satellite system. Such a company would greatly benefit, if it were possible to cluster the data locally at the stores, and then determine and transmit suitable local representatives which allow to reconstruct the complete clustering at the central in Betonville. The transmission of huge amounts of data from one site to another central site is in some application areas almost impossible. In astronomy, for instance, there exist several highly sophisticated space telescopes spread all over the world. These telescopes gather data unceasingly. Each of them is able to collect 1GB of data per hour [5] which can only, with great difficulty, be transmitted to a global site to be analyzed centrally there. On the other hand, it is possible to analyze the data locally where it has been generated and stored. Aggregated information of this locally analyzed data can then be sent to a central site where the information of different local sites are combined and analyzed. The result of the central analysis may be returned to the local sites, so that the local sites are able to put their data into a global context.

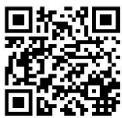

In this paper, we introduce a scalable density-based distributed clustering algorithm which efficiently and effectively detects information spread over several local sites. In our approach, we first compute the density around each locally located object reflecting its suitability to serve as a representative of the local site. After ordering the objects according to their density, we send the most suitable local representatives to a server site, where we cluster the objects by means of an enhanced DBSCAN [4] algorithm. The result is sent back to the local sites. The local sites update their clustering based on the global model, e.g. merge two local clusters to one or assign local noise to global clusters.

This paper is organized as follows: In Section 2, we review the related work in the area of density-based distributed clustering. In Section 3, we discuss a general framework for distributed clustering. In Section 4, we present our quality driven approach for generating local representatives. In Section 5, we show how these representatives can be used for creating a global clustering based on the information transmitted from the local sites. In Section 6, we present the experimental evaluation of our *SDBDC* (Scalable Density-Based Distributed Clustering) approach showing that we achieve high quality clusterings with relative little information. We conclude the paper in Section 7 with a short summary and a few remarks on future work.

## 2 Related Work on Density-Based Distributed Clustering

Distributed Data Mining (DDM) is a dynamically growing area within the broader field of Knowledge Discovery in Databases (KDD). Generally, many algorithms for distributed data mining are based on algorithms which were originally developed for parallel data mining. In [8] some state-of-the-art research results related to DDM are resumed.

One of the main data mining tasks is clustering. There exist many different clustering algorithms based on different paradigms, e.g. density-based versus distance-based algorithms, and hierarchical versus partitioning algorithms. For more details we refer the reader to [7].

To the best of our knowledge, the only density-based distributed clustering algorithm was presented in [6]. The approach presented in [6] is based on the density-based partitioning clustering algorithm DBSCAN. It consists of the following steps. First, a DBSCAN algorithm is carried out on each local site. Based on these local clusterings, cluster representatives are determined. Thereby, the number and type of local representatives is fixed. Only so called special core-points are used as representatives. Based on these local representatives, a standard DBSCAN algorithm is carried out on the global site to reconstruct the distributed clustering. The strong point of [6] is that it tackles the complex and important problem of distributed clustering. Furthermore, it was shown that a global clustering carried out on about 20% of all data points, yields a clustering quality of more than 90% according to the introduced quality measure.

Nevertheless, the approach presented in [6] suffers from three drawbacks which are illustrated in Figure 1 depicting data objects located at 3 different local sites.

- First, local noise is ignored. The clustering carried out on the local sites ignores the local noise located in the upper left corner of each site. Thus the distributed clustering algorithm of [6] does not detect the global cluster in the upper left corner.

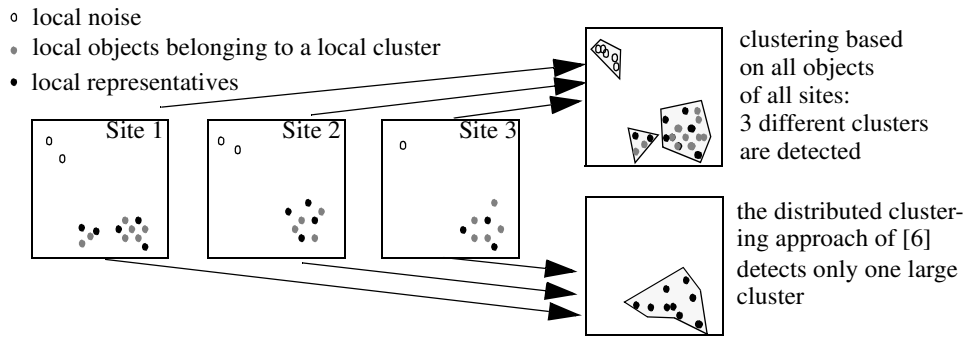

**Fig. 1.** Local noise on different local sites.

- Second, the number of representatives is not tuneable. Representatives are always special core-points of local clusters (cf. black points in Figure 1). The number of these special core-points is determined by the fact that each core point is within the ε-range of a special core-point. Dependent on how the DBSCAN algorithm walks through a cluster, the special core-points are computed.
- Third, these special core points might be located at the border of the clusters (cf. Figure 1). Of course, it is much better if the representatives are not located at the trailing end of a cluster, but are central points of a cluster. Representatives located at the border of local clusters might lead to a false merging of locally detected clusters when carrying out a central clustering. This is especially true, if we use high ε-values for the clustering on the server site, e.g. in [6] a high and static value of $\varepsilon_{global} = 2\varepsilon_{local}$ was used. The bottom right corner of Figure 1 shows that badly located representatives along with high $\varepsilon_{global}$-values might lead to wrongly merged clusters.

To sum up, in the example of Figure 1, for instance, the approach of [6] would only detect one cluster instead of three clusters, because it cannot deal with local noise and tend to merge clusters close to each other. Our new SDBDC approach enhances the approach presented in [6] as follows:

- We deal effectively and efficiently with the problem of local noise.
- We do not produce a fixed number of local representatives, but allow the user to find an individual trade-off between cluster quality and runtime.
- Our representatives reflect dense areas tending to be in the middle of clusters.
- Furthermore, we propose a more effective way to detect the global clustering based on the local representatives. We do not apply a DBSCAN algorithm with a fixed ε-value. Instead we propose to use an enhanced DBSCAN algorithm which uses different ε-values for each local representative *r* depending on the distribution of the objects represented by *r*.

## 3 Scalable Density-Based Distributed Clustering

Distributed Clustering assumes that the objects to be clustered reside on different sites. Instead of transmitting all objects to a central site (also denoted as server) where we can apply standard clustering algorithms to analyze the data, the data is analyzed independently on the different local sites (also denoted as clients). In a subsequent step, the central site tries to establish a global clustering based on the local models, i.e. the local representatives. In contrast to a central clustering based on the complete dataset, the central clustering based on the local representatives can be carried out much faster.

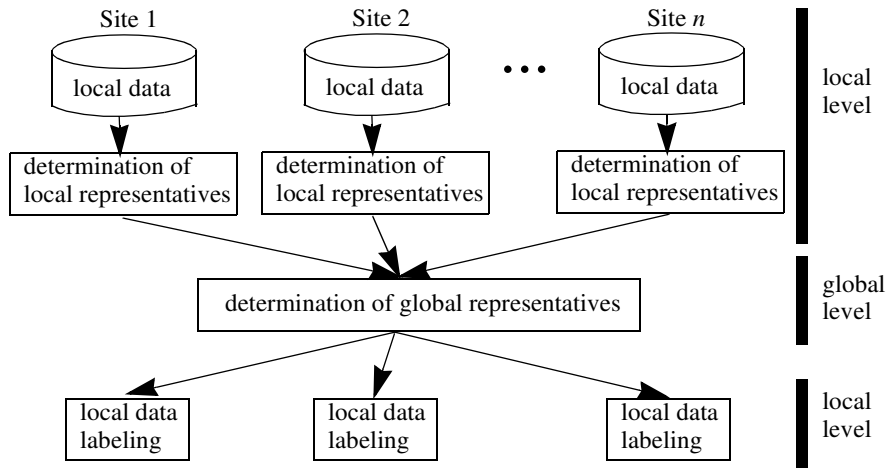

**Fig. 2.** Distributed clustering.

Distributed Clustering is carried out on two different levels, i.e. the local level and the global level (cf. Figure 2). On the local level, all sites analyse the data independently from each other resulting in a local model which should reflect an optimum trade-off between complexity and accuracy. Our proposed local models consist of a set of representatives. Each representative is a concrete object from the objects located at the local site. Furthermore, we augment each representative $r$ with a suitable covering radius indicating the area represented by $r$. Thus, $r$ is a good approximation for all objects residing on the corresponding local sites and are contained in the covering area of $r$.

Next, the local model is transferred to a central site, where the local models are merged in order to form a global model. The global model is created by analysing the local representatives. This analysis is similar to a new clustering of the representatives with suitable global clustering parameters. To each local representative a global cluster identifier is assigned. The resulting global clustering is sent to all local sites.

If a local object is located in the covering area of a global representative, the cluster-identifier from this representative is assigned to the local object. Thus, we can achieve that each site has the same information as if their data were clustered on a global site, together with the data of all the other sites. To sum up, distributed clustering consists of three different steps (cf. Figure 2):

- Determination of a local model
- Determination of a global model which is based on all local models
- Updating of all local models

In this paper, we will present effective and efficient algorithms for carrying out step 1 and step 2. For more details about step 3, the relabeling on the local sites, we refer the interested reader to [6].

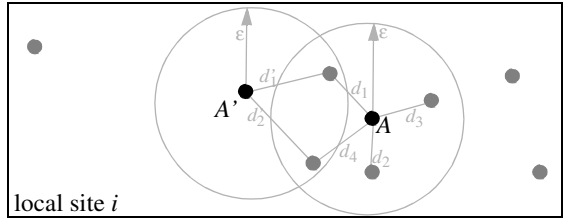

$$StatRepQ(A) = (\varepsilon - d_1) + (\varepsilon - d_2) + (\varepsilon - d_3) + (\varepsilon - d_4) + (\varepsilon - 0)$$
$$>$$
$$StatRepQ(A') = (\varepsilon - d'_1) + (\varepsilon - d'_2) + (\varepsilon - 0)$$

**Fig. 3.** Static representation quality.

## 4 Quality Driven Determination of Local Representatives

In this section, we present a quality driven and scalable algorithm for determining local representatives. Our approach consists of two subsequent steps. First, we introduce and explain the term *static representation quality* which assigns a quality value to each object of a local site reflecting its suitability to serve as a representative. Second, we discuss how the object representation quality changes, dependent on the already determined local representatives. This quality measure is called *dynamic representation quality*. In Section 4.2, we introduce our scalable and quality driven algorithm for determining suitable representatives along with additional aggregated information describing the represented area.

### 4.1 Object Representation Quality

In order to determine suitable local cluster representatives, we first carry out similarity range queries on the local sites around each object $o$ with a radius $\varepsilon$.

**Definition 1** (Similarity Range Query on Local Sites)
Let $O$ be the set of objects to be clustered and $d: O \times O \to IR_0^+$ the underlying distance function reflecting the similarity between two objects. Furthermore, let $O_i \subseteq O$ be the set of objects located at site $i$. For each object $o \in O_i$ and a query range $\varepsilon \in IR_0^+$, the similarity range query $sim_{range}$: $O_i \times IR_0^+ \to 2^{O_i}$ returns the set.

$$sim_{range}(o, \varepsilon) = \{o_i \in O_i \mid d(o_i, o) \leq \varepsilon\}$$

After having carried out the range queries on the local sites, we assign a static representation quality $StatRepQ(o,\varepsilon)$ to each object $o$ w.r.t. a certain $\varepsilon$-value.

**Definition 2** (Static Representation Quality *StatRepQ*)
Let $O_i \subseteq O$ be the set of objects located at site $i$. For each object $o \in O_i$ and a query range $\varepsilon \in IR_0^+$, $StatRepQ$: $O_i \times IR_0^+ \to IR_0^+$ is defined as follows:

$$StatRepQ(o, \varepsilon) = \sum_{o_i \in sim_{range}(o, \varepsilon)} \varepsilon - d(o_i, o)$$

For each object $o_i$ contained in the $\varepsilon$-range of a query object $o$, we determine the distance to the border of the $\varepsilon$-range query, i.e. we weight each object $o_i$ in the $\varepsilon$-range of $o$ by $\varepsilon - d(o_i, o)$. This value is the higher, the closer $o_i$ is to $o$. Then the quality measure $StatRepQ(o,\varepsilon)$ sums up all the values $\varepsilon - d(o_i, o)$ for all objects located in the $\varepsilon$-range of our query object. Obviously, $StatRepQ(o,\varepsilon)$ is the higher, the more objects are located in the $\varepsilon$-range around $o$ and the closer these objects are to $o$. Figure 3 illustrates that the highest $StatRepQ(o,\varepsilon)$ value is assigned to those

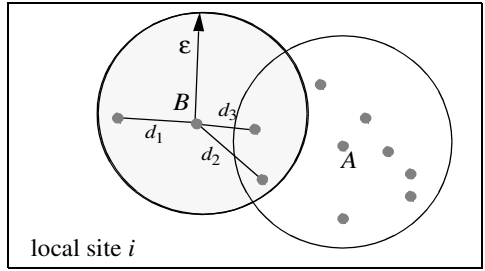

$$DynRepQ(B, \varepsilon, \{ \}) =$$
$$(\varepsilon - d_1) + (\varepsilon - d_2) + (\varepsilon - d_3) + (\varepsilon - 0)$$
$$>$$
$$DynRepQ(B, \varepsilon, \{A\}) =$$
$$(\varepsilon - d_1) + (\varepsilon - 0)$$

**Fig. 4.** Dynamic representation quality.

objects which intuitively seem to be the most suitable representatives of a local site. The figure shows that the value $StatRepQ(A,\varepsilon)$ is much higher than the value $StatRepQ(A',\varepsilon)$, reflecting the more central role of object $A$ compared to object $A'$.

Next we define a dynamic representation quality $DynRepQ(o,\varepsilon,Rep_i)$ for each local object $o$. This quality measure depends on the already determined set of local representatives $Rep_i$ of a site $i$ and the radius of our $\varepsilon$-range query.

**Definition 3** (Dynamic Representation Quality $DynRepQ$)
Let $O_i \subseteq O$ be the set of objects located at site $i$ and $Rep_i \subseteq O_i$ the set of the already determined local representatives of site $i$. Then, $DynRepQ: O_i \times IR_0^+ \times 2^{O_i} \to IR_0^+$ is defined as follows:

$$DynRepQ(o, \varepsilon, Rep_i) = \sum_{\substack{o_i \in sim_{range}(o, \varepsilon) \\ \forall r \in Rep_i: o_i \notin sim_{range}(r, \varepsilon)}} \varepsilon - d(o_i, o)$$

$DynRepQ(o,\varepsilon,Rep_i)$ depends on the number and distances of the elements found in the $\varepsilon$-range of an object $o$, which are not yet contained in the $\varepsilon$-range of a former local representative. For each object $o$ which has not yet been selected as a representative, the value $DynRepQ(o,\varepsilon,Rep_i)$ gradually decreases with an increasing set of local representatives, i.e. an increasing set $Rep_i$. Figure 4 illustrates the different values of $DynRepQ(B,\varepsilon,Rep_i)$ for two values of the set $Rep_i$. If $Rep_i=\{\}$, the value $DynRepQ(B,\varepsilon,Rep_i)$ is much higher than if the element $A$ is included in $Rep_i$.

### 4.2 Scalable Calculation of Local Representatives

In this subsection, we will show how we can use the quality measures introduced in the last subsection to create a very effective and efficient algorithm for determining a set of suitable local representatives. The basic idea of our greedy algorithm is very intuitive (cf. Figure 5).

- First, we carry out range queries for each object of a local site.
- Second, we sort the objects in descending order according to their static representation quality.
- Third, we delete the first element from the sorted list and add it to the set of local representatives.
- Fourth, we compute the dynamic representation quality for each local object which has not yet been used as a local representative and sort these objects in descending order according to their dynamic representation quality.
- If we have not yet determined enough representatives, we continue our algorithm with step 3. Otherwise, the algorithm stops.

```
O_i         set of objects located at site i;
ε           ε-range value;
ALGORITHM DeterminationOfLocalRepresentatives;
BEGIN
    Rep_i :={ };                                              // set of local representatives;
    FOR EACH o ∈ O_i DO
        compute StatRepQ(o,ε);
    END FOR;
    SortRepList := <(o_1,StatRepQ(o_1,ε)), ..., (o_|O_i|,StatRepQ(o_|O_i|,ε))| i≤j =>StatRepQ(o_i,ε) ≥StatRepQ(o_j,ε)>;
    WHILE NOT stop_criterion (Rep_i) DO
        Rep_i := Rep_i + SortRepList[1];
        FOR EACH o ∈ O_i − REP_i DO
            compute DynRepQ(o,ε,Rep_i);
        END FOR;
        SortRepList := <(o_1,DynRepQ(o_1,ε,Rep_i)), ..., (o_|O_i-REP_i|,DynRepQ(o_|O_i-REP_i|,ε,Rep_i))|
                        i ≤ j =>DynRepQ(o_i,ε,Rep_i) ≥ DynRepQ(o_j,ε,Rep_i)>;
    END WHILE;
END.
```

**Fig. 5.** Scalable calculation of local representatives.

Obviously, the algorithm delivers the most suitable local representatives at a very early stage of the algorithm. After having determined a new local representative, it can be sent to a global site without waiting for the computation of the next found representative. As we decided to apply a greedy algorithm for the computation of our local representatives, we will not revoke a representative at a later stage of the algorithm. So the algorithm works quite similar to ranking similarity queries known from database systems allowing to apply the cursor principle on the server site. If the server decides that it has received enough representatives from a local site, it can close the cursor, i.e. we do not have to determine more local representatives. The termination of the algorithm can either be determined by a size-bound or an error-bound stop criterion [10]. This approach is especially useful if we apply a clustering algorithm on the server site which efficiently supports incremental clustering as, for instance, DBSCAN [3].

For all representatives included in a sequence of local representatives, we also compute their Covering Radius *CovRad*, indicating the element which has the maximum distance from the representative, and the number *CovCnt* of objects covered by the representative.

**Definition 4** (Covering Radius and Covering Number of Local Representatives)
Let $O_i \subseteq O$ be the set of objects located at site $i$ and $Rep_{i_n} = \{r_{i_1}, ..., r_{i_n}\}$ the sequence of the first $n$ local representatives where $\{r_{i_1}, ..., r_{i_n}\} \subseteq O_i$. Then the covering radius *CovRad*: $O_i \times IR_0^+ \times 2^{O_i} \to IR_0^+$ and the covering number *CovCnt*: $O_i \times IR_0^+ \times 2^{O_i} \to IR_0^+$ of the $i_{n+1}$th representative are defined as follows:

$$CovRad(r_{i_{n+1}}, \varepsilon, Rep_{i_n}) =$$
$$max\{\varepsilon - d(o, r_{i_{n+1}}) | \forall o \in O_i \forall r \in Rep_{i_n} : o \in sim_{range}(r_{i_{n+1}}, \varepsilon) \land o \notin sim_{range}(r, \varepsilon)\}$$

$$CovCnt(r_{i_{n+1}}, \varepsilon, Rep_{i_n}) = \left|\{o | \forall o \in O_i \forall r \in Rep_{i_n} : o \in sim_{range}(r_{i_{n+1}}, \varepsilon) \land o \notin sim_{range}(r, \varepsilon)\}\right|$$

Figure 6 depicts the *CovRad* and *CovCnt* values for two different representatives of site *i*. Note that the computation of *CovRad* and *CovCnt* can easily be integrated into the computation of the representatives as illustrated in Figure 5. The local representatives along with the corresponding values *CovRad* and *CovCnt* are sent to the global site in order to reconstruct the global clustering, i.e. we transmit the following sequence consisting of *n* local representatives from site *i*:

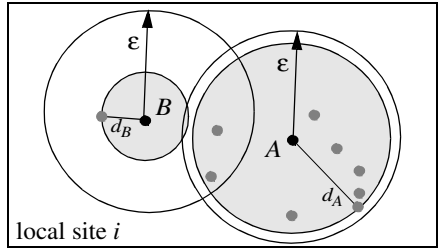

$$CovRad(A, \langle \rangle) = d_A$$
$$CovCnt(A, \langle \rangle) = 9$$
$$CovRad(B, \langle A \rangle) = d_B$$
$$CovCnt(B, \langle A \rangle) = 2$$

sent to global site: $< (A, d_A, 9),\ (B, d_B, 2)>$

**Fig. 6.** Covering radius *CovRad* and covering number *CovCnt*.

$$< (r_{i_1}, CovRad(r_{i_1}, \varepsilon, \{\}), CovCnt(r_{i_1}, \varepsilon, \{\})),$$
$$\vdots$$
$$(r_{i_n}, CovRad(r_{i_n}, \varepsilon, \{r_{i_1},..,r_{i_{n-1}}\}), CovCnt(r_{i_n}, \varepsilon, \{r_{i_1},..,r_{i_{n-1}}\})) >.$$

For simplicity, we will write $CovRad(r_{i_j})$ instead of $CovRad(r_{i_j}, \varepsilon, \{r_{i_1},..,r_{i_{j-1}}\})$ and $CovCnt(r_{i_j})$ instead of $CovCnt(r_{i_j}, \varepsilon, \{r_{i_1},..,r_{i_{j-1}}\})$, if the set of already transmitted representatives and the used ε-values are clear from the context.

## 5 Global Clustering

On the global site, we apply an enhanced version of DBSCAN adapted to clustering effectively local representatives. We carry out ε-range queries around each representative. Thereby, we use a specific ε-value $\varepsilon(r_i)$ for each representative $r_i$ (cf. Figure 7). The ε-value $\varepsilon(r_i)$ is equal to the sum of the following two components. The first component consists of the basic ε-value which would be used by the original DBSCAN algorithm and which was used for the range queries on the local sites. The second component consists of the specific $CovRad(r_i)$ value of the representative $r_i$, i.e we set $\varepsilon(r_i) = \varepsilon + CovRad(r_i)$.

The idea of this approach is as follows (cf. Figure 8). The original DBSCAN algorithm based on all data of all local sites would carry out an ε-range query around each point of the data set. As we perform the distributed clustering only on a small fraction of these weighted points, i.e. we cluster on the set of the local representatives transmitted from the different sites, we have to enlarge the ε-value by the $CovRad(r_i)$ value of the actual representative $r_i$. This approach guarantees that we can find all objects in the enlarged ε-range query which would have been found by any object represented by the actual local representative $r_i$. For instance in Figure 7, the representative $r_{j_1}$ is within the ε-range of the local object $o_i$ represented by $r_{i_1}$. Only because we use the enlarged ε-range $\varepsilon(r_{i_1})$, we detect that the two representatives $r_{i_1}$ and $r_{j_1}$ belong to the same cluster.

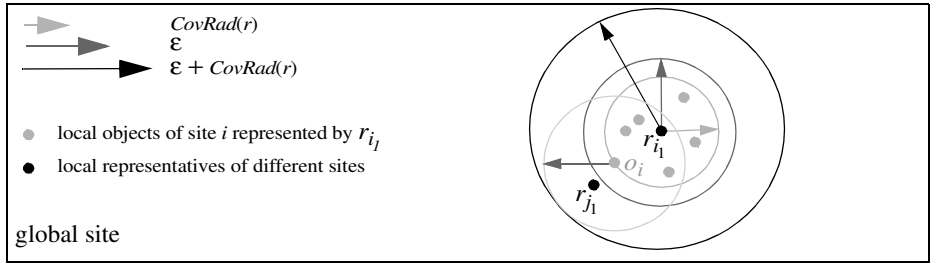

**Fig. 7.** Global clustering on varying $\varepsilon(r_i)$-parameters for the different representatives $r_i$

Furthermore, we weight each local representative $r_i$ by its $CovCnt(r_i)$ value, i.e. by the number of local objects which are represented by $r_i$. By taking these weights into account, we can detect whether local representatives are core-points, i.e. points which have more than *MinPts* other objects in their ε-range. For each core-point $r_j$ contained in a cluster $C$, we carry out an enlarged ε-range query with radius $\varepsilon(r_j)$ trying to expand $C$. In our case, a local representative $r_j$ might be a core-point although less than *MinPts* other local representatives are contained in its $\varepsilon(r_j)$-range. For deciding whether $r_j$ is a core-point, we have to add up the number of objects $CovCnt(r_i)$ represented by the local representatives $r_i$ contained in the $\varepsilon(r_j)$-range of $r_j$ (cf. Figure 8).

```
R                set of all representatives from all local sites;
                 // R ={(r1, CovRad(r1), CovCnt(r1)), .., (r|R|, CovRad(r|R|), CovCnt(r|R|));
ε, MinPts        ε-range value and MinPts parameter used by DBSCAN
Algorithm DistributedGlobalDBSCAN
BEGIN
    ActClusterId := 1;                                       // ClusterId = 0 is used for NOISE and
    FOR i =1 .. |R| DO                                       // ClusterId = -1 for UNCLASSIFIED objects
        ActObj := R.get(i);                                  // select the ith object from R
        IF ActObj.ClusterId = -1 THEN
            IF ExpandCluster THEN
                ActClusterId:=ActClusterId +1;
            END IF;
        END IF;
    END FOR;
END.
ExpandCluster: Boolean;
BEGIN
    seeds := RangeQuery(ActObj, ε+CovRad(ActObj));           // range query with enlarged radius around ActObj
    CntObjects := 0;
    FOR i = 1 .. |seeds| DO
        CntObjects := CntObjects + seeds[i].CovCnt;          // all objects represented by representatives
    END FOR;
    IF CntObjects < MinPts THEN                              // Object ActObj is not a core object
        ActObj.ClusterID := 0                                // ClusterID 0 is used for NOISE
        RETURN FALSE;
    ELSE                                                     // Object o is a core object
        FOR i = 1 .. |seeds| DO
            IF seeds[i].ClusterId = {-1, 0} THEN
                seeds[i].ClusterId := ActClusterId;
            END IF;
        END FOR;
        delete ActObj from seeds;
        WHILE seeds NOT EMPTY DO
            ActObj := seeds[1];
            neighborhood := RangeQuery(ActObj, ε+CovRad(ActObj));   // range query with enlarged radius
            CntObjects := 0
            FOR i = 1 .. |neighborhood| DO
                CntObjects := CntObjects + neighborhood[i].CovCnt;
            END FOR;
            IF CntObjects >= MinPts THEN                     // ActObj is a core object
                FOR i = 1 .. |neighborhood| DO
                    p := neighborhood[i];
                    IF p.ClusterId = {-1, 0} THEN            // object p is UNCLASSIFIED or NOISE
                        IF p.ClusterId = -1 THEN             // object p is UNCLASSIFIED
                            add p to seeds;
                        END IF;
                        p.ClusterId := ActClusterID;
                    END IF;
                END FOR;
            END IF;
            delete ActObj from seeds;
        END WHILE;
        RETURN TRUE;
    END IF;
END;
```

**Fig. 8.** Distributed global DBSCAN algorithm.

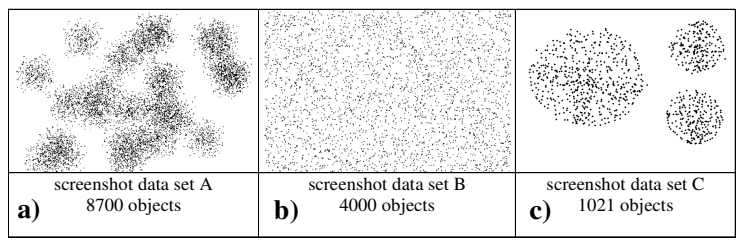

**Fig. 9.** Used test data sets.
**a)** test data set *A,* **b)** test data set *B,* **c)** test data set *C*

## 6 Experimental Evaluation

We evaluated our SDBDC approach based on three different 2-dimensional point sets where we varied both the number of points and the characteristics of the point sets. Figure 9 depicts the three used test data sets *A* (8700 objects, randomly generated data/clusters), *B* (4000 objects, very noisy data) and *C* (1021 objects, 3 clusters) on the central site.

In order to evaluate our SDBDC approach, we equally distributed the data set onto the different client sites and then compared SDBDC to a single run of DBSCAN on all data points. We carried out all local clusterings sequentially. Then, we collected all representatives of all local runs, and applied a global clustering on these representatives. For all these steps, we used a Pentium III/700 machine. In all experiments, we measured the overall number of transmitted local representatives, which primarily influences the overall runtime. Furthermore, we measured the cpu-time needed for the distributed clustering consisting of the maximum time needed for the local clusterings and the time needed for the global clustering based on the transmitted local representatives.

We measured the quality of our SDBDC approach by the quality measure introduced in [6]. Furthermore we compared our approach to the approach presented in [6] where for the three test data sets about 17% of all local objects were used as representatives. Note that this number is fixed and does not adapt to the requirements of different users, i.e high clustering quality or low runtime.

Figure 10 shows the trade-off between the clustering quality and the time needed for carrying out the distributed clustering based on 4 different local sites.

Figure 10a shows clearly that with an increasing number of local representatives the overall clustering quality increases. For the two rather noisy test data sets *A* and *B* reflecting real-world application ranges, we only have to use about 5% of all local objects as representatives in order to achieve the same clustering quality as the one achieved by the approach presented in [6].

Figure $10b_1$ shows the speed up w.r.t. the transmission cost we achieve when transmitting only the representatives determined by our SDBDC approach compared to the transmission of all data from the local sites to a global site. We assume that a local object is represented by *n* bytes and that both *CovRad*($r_i$) and *CovCnt*($r_i$) need about 4 bytes each. For realistic values of *n*, e.g. *n*=100, a more than three times lower representative number, e.g. 5% used by the SDBDC approach compared to 17% used by the approach presented in [6], results in a 300% speed up w.r.t. the overall transmission cost which dominate the overall runtime cost (cf. Figure $10b_1$).

Figure $10b_2$ depicts the sum of the maximum cpu-time needed for the clustering on the local site and the cpu-time needed for the clustering on the global site. A small number of local representatives terminates the generation of the local representatives at an early stage leading to a short runtime for computing the required local representatives. Furthermore, the global clustering can be carried out the more efficiently, the smaller the overall number of local representatives is.

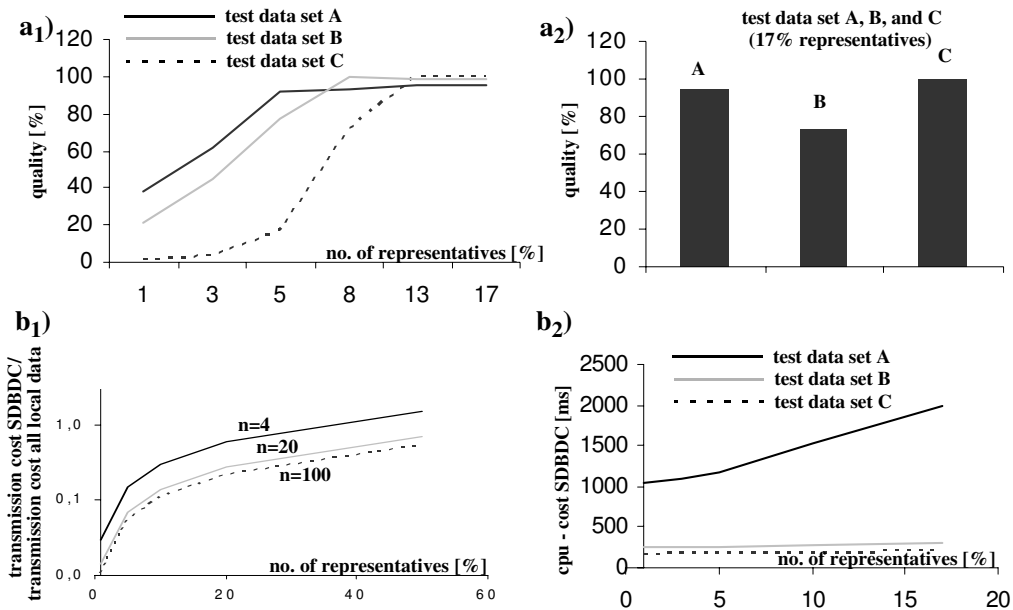

**Fig. 10.** Trade-off between runtime and clustering quality (4 sites).
**a)** clustering quality (**a₁**) SDBDC **a₂**) approach presented in [6]),
**b)** runtime (**b₁**) transmission cost **b₂**) cpu-cost for local and global clustering)

To sum up, a small number of local representatives accelerates the SDBDC approach considerably. If we use about 5% of all objects as representatives, we can achieve a high quality and, nevertheless, efficient distributed clustering.

Figure 11 shows how the clustering quality depends on the number of local sites. Obviously, the quality decreases when increasing the number of sites. This is especially true for the approach presented in [6] which neglects the problem of local noise. The more sites we have and the noisier the data set is, the more severe this problem is. As Figure 11 shows, our approach is much less susceptible to an increasing number of local sites. Even for the noisy test data set *B*, our approach stays above 90% clustering quality although using more than 10 different local sites and only 13% of all objects as local representatives (in contrast to the fixed 17% used in [6]).

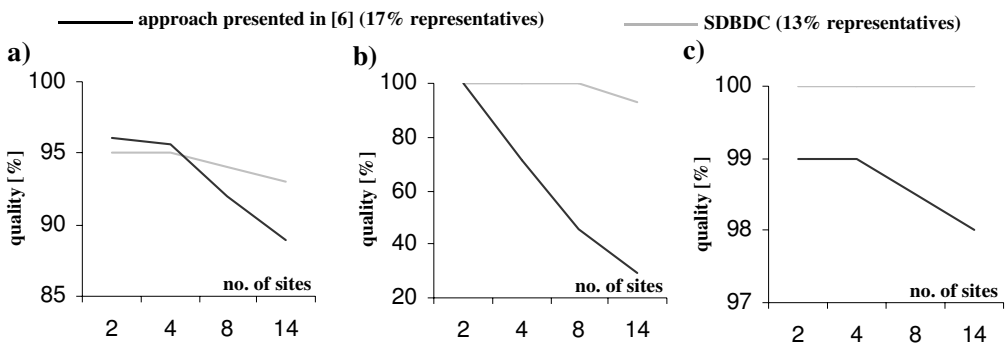

**Fig. 11.** Clustering Quality dependent on the number of local sites.
**a)** test data set *A,* **b)** test data set *B,* **c)** test data set *C*

# 7 Conclusions

In this paper, we first discussed some application ranges which benefit from an effective and efficient distributed clustering algorithm. Due to economical, technical and security reasons, it is often not possible to transmit all data from different local sites to one central server site where the data can be analysed by means of clustering. Therefore, we introduced an algorithm which allows the user to find an individual trade-off between clustering-quality and runtime. Our approach first analyses the data on the local sites and orders all objects $o$ according to a quality criterion *DynRepQ(o)* reflecting whether the actual object is a suitable representative. Note that this quality measure depends on the already determined representatives of a local site. After having transmitted a user dependent number of representatives to the server, we apply a slightly enhanced DBSCAN clustering algorithm which takes the covering radius and the number of objects covered by each representative $r_i$ into account, i.e. the server site clustering is based on the aggregated information *CovRad*($r_i$) and *CovCnt*($r_i$) describing the area on a local site around a representative $r_i$. As we produce the local representatives in a give-me-more manner and apply a global clustering algorithm which supports efficient incremental clustering, our approach allows to start with the global clustering algorithm as soon as the first representatives are transmitted from the various local sites. Our experimental evaluation showed that the presented scalable density-based distributed clustering algorithm allows effective clustering based on relatively little information, i.e. without sacrificing efficiency and security.

In our future work, we plan to develop hierarchical distributed clustering algorithms which are suitable for handling nested data.